\documentstyle[aps,prl,floats,graphicx,epsfig]{revtex}
\begin{document}
\draft

\twocolumn[\hsize\textwidth\columnwidth\hsize\csname @twocolumnfalse\endcsname
\title{Rotational quantum friction in superfluids: Radiation from object
rotating
in superfluid vacuum.}
\author{A Calogeracos$^{1,2}$ and G.E. Volovik$^{1,3}$}
\address{
$^{1}$Low Temperature Laboratory,
Helsinki University of Technology,
P.O.Box 2200, FIN-02015, Finland\\
$^{2}$ NCA\ Research Associates,
 PO\ Box 61147, Maroussi 151 01,
Athens, Greece \\
$^{3}$ L.D. Landau Institute for Theoretical Physics,
Kosygin Str. 2, 117940 Moscow, Russia\\}
\date{\today}
\maketitle

\begin{abstract}

{We discuss the friction experienced by the body rotating in superfluid
liquid at
$T=0$. The effect is analogous to the amplification of electromagnetic
radiation
and spontaneous emission by the body or black hole rotating in quantum
vacuum, first
discussed by Zel'dovich and Starobinsky. The friction is caused by the
interaction
of the part of the liquid, which is rigidly connected with the rotating
body and
thus represents the comoving detector, with the "Minkowski"  vacuum outside
the body. The emission process is the quantum tunneling of quasiparticles
from the detector to the ergoregion, where the energy of quasiparticles  is
negative in the rotating frame. This quantum rotational friction caused by
the emission of quasiparticles is estimated for phonons and rotons in
superfluid $^4$He and for  Bogoliubov fermions in superfluid $^3$He.}
\end{abstract}

\

\pacs{PACS numbers:  67.40.-w, 67.57.Fg, 04.70.-s, 42.50.-p}

\

] \narrowtext

\vfill\eject
\subsection{Introduction.}

The  body moving in the vacuum  with linear  acceleration $a$ is believed to
radiate the thermal spectrum with the Unruh temperature $T_U=\hbar a/2\pi c$
\cite{Unruh1}. The comoving observer sees the vacuum as a thermal bath
with $T=T_U$, so that the matter of the body gets heated to $T_U$ (see
references
in \cite{Audretsch}). Linear motion at constant proper acceleration (hyperbolic
motion) leads to arbitrary high velocity. On the other hand uniform
circular motion
features constant centripetal acceleration while being free of the pathology of
infinite velocity (see the latest references in
\cite{RotatingQuantumVacuum,Leinaas,OrbitingUnruh}). The latter motion is
stationary in the rotating frame, which is thus a convenient  frame for
study of the radiation and thermalization effects for uniformly rotating
body.

Zel'dovich \cite{Zeldovich1} was the first who predicted that the rotating
body (say, dielectric cylinder) amplifies those electromagnetic modes which
satisfy the condition
\begin{equation}
 \omega - L \Omega<0 ~.
\label{ZeldovichCondition}
\end{equation}
Here $\omega$ is the frequency of the mode, $L$ is its azimuthal quantum
number, and $\Omega$ is the angular velocity of the rotating cylinder.
This amplification of the incoming radiation is referred to as
superradiance
\cite{BekensteinSchiffer}. The other aspect of this phenomenon is that due
to quantum effects, the cylinder rotating in quantum vacuum
spontaneously emits the electromagnetic modes satisfying
Eq.(\ref{ZeldovichCondition})
\cite{Zeldovich1}. The same occurs for any rotating body, including the
rotating black hole \cite{Starobinskii}, if the above condition is satisfied.

Distinct from the linearly accelerated body, the radiation by a rotating
body does not look thermal. Also, the rotating observer does not see the
Minkowski vacuum as a thermal bath.  This means that the matter of the body,
though excited by interaction with the quantum fluctuations of the Minkowski
vacuum, does not necessarily acquire an intrinsic temperature  depending
only on
the angular velocity of rotation. Moreover the vacuum of the rotating frame
is not
well defined because of the ergoregion, which exists at the distance
$r_e=c/\Omega$ from the axis of rotation.

The problems related to the response of the quantum system  in its ground
state to rotation\cite{RotatingQuantumVacuum}, such as radiation by the
object rotating in vacuum
\cite{Zeldovich1,Zeldovich2,Starobinskii,BekensteinSchiffer} and  the
vacuum instability caused by the existense of ergoregion
\cite{QuantumErgoregionInstablity}, etc., can be simulated in superfluids,
where
the superfluid ground state plays the part of the quantum vacuum. We
discuss the
quantum friction due to spontaneous emission of phonons and rotons
in superfluid $^4$He and Bogoliubov fermions in superfluid $^3$He.

\subsection{Rotating frame.}

Let us consider a cylinder of radius $R$ rotating with angular
velocity $\Omega$ in the (infinite) superfluid liquid. In bosonic
superfluids the
quasiparticles are phonons and rotons; in fermi superfluids these are the
Bogoliubov fermions. The phonons are "relativistic" quasiparticles: Their
energy spectrum is $E(p)=cp+\vec p\cdot \vec v_s$, where $c$ is a speed of
sound and
$\vec v_s$ is the superfluid velocity, the velocity of the superfluid
vacuum; and
this phonon  dispersion is represented by the Lorentzian metric (the
so-called acoustic metric \cite{UnruhSonic,Visser}):
 \begin{equation}
g^{\mu\nu}p_\mu p_\nu=0~~,~~g^{00}=-1, g^{0i}=v_s^i~,
 g^{ik}=c^2\delta^{ik}-v_s^iv_s^k
 ~.
\label{Metric}
\end{equation}

When the body rotates, the
energy of quasiparticles is not well determined in the laboratory frame due
to the
time dependence of the potential, caused by the rotation of the body. But it is
determined in the rotating frame, where the potential is stationary. Hence it
is simpler to work in the rotating frame. If the body is rotating
surrounded by the
stationary superfluid, i.e.
$\vec v_s=0$ in the laboratory frame, then in the rotating frame one has $\vec
v_s=-\vec\Omega\times
\vec r$. Substituting this $\vec v_s$ in Eq.(\ref{Metric}) we get  the interval
$ds^2=g_{\mu\nu}dx^\mu dx^\nu$, which determines the propagation of phonons
in the
rotating frame:
 \begin{equation}
ds^2=-(c^2-\Omega^2r^2)dt^2 - 2\Omega r^2d\phi dt +dz^2 + r^2d\phi^2+dr^2
 ~.
\label{Interval}
\end{equation}
The azimuthal motion of the quasiparticles in the rotating frame can be
quantized in
terms of the angular momentum
$L$, while the radial motion can be treated in the quasiclassical
approximation.
Then the energy spectrum of the phonons in the rotating frame is
\begin{equation}
E = c\sqrt{ {L^2\over r^2} +
p_z^2 +  p_r^2} - \Omega L
 ~.
\label{QuasiclassicalPhononSpectrum}
\end{equation}

\subsection{Ergoregion in superfluids.}

The radius $r_e=c/\Omega$, where $g_{00}=0$, marks the position of
the ergoplane. In the ergoregion, i.e. at
$r>r_e=c/\Omega$, the energy of quasiparticle
in Eq.(\ref{QuasiclassicalPhononSpectrum}) can become negative for any
rotation velocity and $\Omega L> 0$. We assume that the angular velocity of
rotation $\Omega$ is small enough, so that the linear velocity on the
surface of the cylinder $\Omega R$ is less than $v_L=c$ (the Landau
velocity for nucleation of phonons). Thus phonons cannot be
nucleated at the surface of cylinder. However in the ergoplane the velocity
$v_s=\Omega r$ in the rotating frame reaches $c$, so that quasiparticle can
be created in the ergoregion $r>r_e$.

The process of creation is, however, determined by the dynamics, i.e. by the
interaction with the rotating body; there is no radiation in the absence of the
body. If $\Omega R\ll v_L=c$ one has $r_e\gg R$, i.e.   the ergoregion is
situated
far from the cylinder; thus the interaction of the phonons state in the
ergoregion with the rotating body is small. This results in a small
emission rate
 and thus in a small value of quantum friction, as will be discussed below.

Let us now consider other excitations: rotons and Bogoliubov fermions. Their
spectra in the rotating frame are
\begin{eqnarray}
E(p)=\Delta + {(p-p_0)^2\over 2m_0} -\Omega L~,\\
E(p)=\sqrt{\Delta^2 +
v_F^2(p-p_0)^2} -\Omega L
 ~.
\label{RotonBogolonSpectrum}
\end{eqnarray}
Here $p_0$ marks the roton minimum in superfluid $^4$He and the Fermi
momentum in
Fermi liquid, while $\Delta$ is either a roton gap or   the gap in
superfluid $^3$He-B. The Landau critical velocity for the emission of these
quasiparticles is
$v_{L}=min~ {E(p)\over p}\sim
\Delta/p_0$.  In $^4$He the Landau velocity for emission of rotons is
smaller than
that for the emission of phonons,
$v_{L}=c$. That is why the ergoplane for rotons, $r_e=v_L/\Omega$, is
closer to the
cylinder.  However, for the rotating body the emission of the rotons is
exponentially suppressed due to the big value of the allowed angular
momentum for
emitted rotons: the Zel'dovich condition Eq.(\ref{ZeldovichCondition}) for
roton
spectrum is satified only for
$L>\Delta/\Omega \gg 1$  (see Fig. \ref{TunnelingToErgoregion}b).

\subsection{Rotating detector.}

Let us consider the system, which is rigidly connected to
the rotating body and thus comprises the comoving detector. In superfluids the
simplest model for such a detector consists of the layer near the surface of
the cylinder, where the superfluid velocity follows the rotation of
cylinder, i.e. $\vec v_s=\vec\Omega\times \vec r$ in the laboratory frame
and thus
$\vec v_s=0$ in the rotating frame. This means that, as distinct from the
superfluid outside the cylinder, in such a layer the quasiparticle spectrum
has no
$-\Omega L$ shift of the energy levels.

Since in the detector matter, i.e. in the surface layer, the vorticity in the
laboratory frame is nonzero,
$\vec\nabla\times
\vec v_s=2\vec\Omega\neq 0$, this layer either contains vortices or
is represented by the normal (nonsuperfluid) liquid, which is rigidly
rotating with
the body.
Actually the whole rotating cylinder can be represented by the rotating normal
liquid.   The equilibrium
state of the rotating normal liquid, viewed in the rotating frame, is the
same as the
equilibrium stationary normal liquid, viewed in the laboratory frame. The
rotating
cylinder can also be represented  by the cluster of the quantized
vortices. Such rigidly rotating clusters of vortices are experimentally
investigated in superfluid
$^3$He (see e.g. \cite{Cluster}).

\begin{figure}[!!!t]
\begin{center}
\leavevmode
\epsfig{file=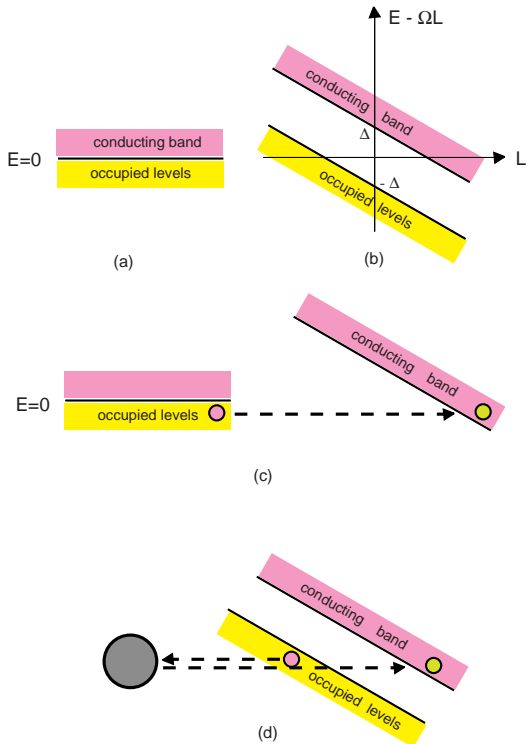,width=0.8\linewidth}
\caption[TunnelingToErgoregion]
    {(a) Vacuum of the Fermi liquid within the rotating body at $r<R$. This
vacuum is rotating together with the body and thus plays the role of the
comoving detector. (b) "Minkowski" vacuum of superfluid outside the
rotating body as viewed in rotating frame. In the ergoregion, i.e. at
$r>r_e=v_L/\Omega$, where $v_L$ is Landau critical velocity,  the
conducting band crosses the zero energy level.  (c) Tunneling  of particles
from the vacuum of the detector matter to the "Minkowski" vacuum in the
ergoregion produces radiation from rotating body and excitation of comoving
detector. (d) Transition between the states in the "Minkowski" vacuum due
to interaction with the rotating detector. }
\label{TunnelingToErgoregion}
\end{center}
\end{figure}

We can discuss the complete system as consisting of two parts, each in
its own ground state (see Figs.
\ref{TunnelingToErgoregion}(a-b) for the case of Fermi liquid): (1) The
matter of the detector in its ground state as seen in the rotating frame;
(2) The superfluid outside the cylinder  in its ground state (the
"Minkowski" vacuum) in the laboratory frame. The radiation of
fermions by the rotating cylinder
is described  by the rotating observer as  a
tunneling process (Fig.
\ref{TunnelingToErgoregion}c): fermions tunnel from the occupied
negative energy levels in the detector  to the unoccupied negative energy
state in the ergoregion. The same can be considered as the
spontaneous nucleation of pairs: the particle is nucleated in the
ergoregion and its partner hole is nucleated in the comoving detector. This
process causes the radiation from the rotating body and also the excitation
of the detector.  From the point of view of the Minkowski
(stationary) observer this is described as the excitation of the superfluid
system by the time dependent perturbations.

\subsection{Radiation of phonons to the ergoregion.}

For the Bose case  the radiation of
phonons can be also considered as the process in which the particle in the
normal Bose liquid in the detector tunnels to the scattering state at the
ergoplane,
where also the energy is
$E=0$. In the quasiclassical approximation the tunneling probability is
$e^{-2S}$, where at $p_z=0$:
\begin{equation}
S={\rm {Im}} \int dr~p_r=L \int^{r_e}_{R}dr~\sqrt{{1\over r^2}-  {1\over
r_e^2}}\approx  L  \ln {r_e\over R}
 ~.
\label{Tunneling}
\end{equation}
Thus all the particles with
$L>0$ are radiated, but the radiation probability decreases at higher $L$.
If the
linear velocity at the surface is much less than the Landau critical velocity
$\Omega R \ll c$, the probability of radiation of phonons with the energy
(frequency)
$\omega=\Omega L$  is
\begin{equation}
w \propto e^{-2S}=\left({R\over r_e}\right)^{2L}=\left({\Omega R\over
c}\right)^{2L} = \left({\omega R\over
cL}\right)^{2L}~,~\Omega R \ll c
 ~.
\label{PhononTunnelingProbability}
\end{equation}
If $c$ is substituted by the speed of light,
Eq.(\ref{PhononTunnelingProbability}) is proportional to the superradiant
amplification of the electromagnetic  waves by rotating  dielectric
cylinder derived by Zel'dovich \cite{BekensteinSchiffer,Zeldovich2}.

The number of phonons  with the  frequency
$\omega=\Omega L$  emitted per unit time can be estimated as
$
\dot N= W e^{-2S}
$,
where $W$ is the attempt frequency $\sim \hbar/ma^2$ multiplied by the number
of localized modes $\sim RZ/a^2$, where $Z$ is the height of the cylinder.
Since each phonon carries the angular momentum $L$, the cylinder  rotating in
superfluid vacuum (at $T=0$) is loosing its angular momentum, which means
the quantum rotational friction.

\subsection{Radiation of rotons and Bogoliubov quasiparticles.}

The minimal $L$ value of the radiated quasiparticles, which have the gap
$\Delta$, is
determined by this gap:
$L_{min}={\Delta\over \Omega p_0}={v_L\over \Omega}$, where
$v_L=\Delta/p_0$ is the Landau critical velocity. Since the tunneling rate
exponentially decreases with
$L$, only the lowest possible $L$ must be considered. In this case the
tunneling
trajectory with
$E=0$ is determined by the equation
$p=p_0$ both for rotons and Bogoliubov quaiparticles. For $p_z=0$ the classical
tunneling trajectory is thus given by
$p_r=i\sqrt{|p_0^2-L^2/r^2|}$. This gives for the tunneling exponent
$e^{-2S}$ the equation
\begin{equation}
S={\rm {Im}} \int dr~p_r=L \int^{r_e}_{R}dr~\sqrt{{1\over r^2}-
{1\over r_e^2}}\approx  L  \ln {r_e\over R}
 ~.
\label{RotonTunnelingAction}
\end{equation}
Here the position of the ergoplane is $r_e=L/p_0= v_L/\Omega$.
Since the rotation velocity $\Omega$ is always much smaller than the gap,
$L$ is very
big. That is why the radiation of rotons and Bogoliubov quasiparticles with the
gap is exponentially suppressed.

\subsection{Friction due to transitions in "Minkowski vacuum".}

Radiation can occur without excitation of the
 detector vacuum, via direct interaction of the particles in the Minkowski
vacuum with the rotating body. In the rotating frame the states in the
occupied band and in the conducting band have the same energy, if they have
opposite momenta $L$. Then a transition between the two levels is
energetically allowed and will occur if the Hamiltonian has a nonzero
matrix element between the states
$L$ and
$-L$. The necessary interaction is provided by any
violation of the axial symmetry of the rotating body, e.g. by roughness on
the surface (thus the interaction is localized ar $r\sim R$). A wire moving
along the circular orbit is another practical example. In case of the
rotating vortex cluster the axial symmetry is always violated.

In the quasiclassical approximation the process of radiation is as follows.
The particle from the occupied band in the ergoregion tunnels to the
surface of the rotating body, where after interaction with the
nonaxisymmetric disturbance it changes its angular momentum. After that it
tunnels back to the ergoregion to the conducting band. In this process both
a particle and a hole are produced in the Minkowski vacuum, as a result
the tunneling exponent is twice larger than in
Eqs.(\ref{PhononTunnelingProbability}) and (\ref{RotonTunnelingAction}).

\subsection{Discussion.}

The rotational friction experienced by the body rotating in superfluid
vacuum at
$T=0$, is caused by the  spontaneous quantum emission of the quasiparticles
from
the rotating object to the "Minkowski" vacuum in the ergoregion. The
emission is not
thermal and depends on the details of the interaction of the radiation with the
rotating body.  In the quasiclassical approximation it is mainly determined
by the tunneling exponent, which can be approximately characterized by the
effective temperature
$T_{\rm eff}\sim \hbar \Omega (2/\ln (v_L/\Omega R))$. The vacuum friction
of the
rotating body can be observed only if the effective temperature exceeds the
temperature of the bulk superfluid, $T_{\rm eff}>T$. For the body rotating with
$\Omega=10^3$rad/s,
$T$ must be below $10^{-8}$K. However, high rotation velocity can be
obtained in the
system of two like vortices, which rotate around their center of mass
with
$\Omega=\kappa/\pi R^2$ ($\kappa$ is the circulation around each vortex,
$R$ is the
radius of the  circular orbit).

The process discussed in the paper occurs only if there is an ergoplane in the
rotating frame. For the superfluid confined within the external cylinder of
radius
$R_{\rm ext}$, this process occurs at high enough rotation velocity,
$r_e(\Omega)=v_L/\Omega < R_{\rm ext}$, when the ergoplane is
within the superfluid.   On the instability of the ergoregion in
quantum vacuum towards emission see e.g.
Ref.\cite{QuantumErgoregionInstablity}.

If  $r_e(\Omega)> R_{\rm ext}$ and ergoregion is not present, then the
interaction between the coaxial cylinders via the vacuum
fluctuations becomes the main mechanism for dissipation. This causes the
dynamic Casimir forces between the walls moving laterally  (see Review
\cite{Kardar}). As in \cite{Kardar} the nonideality of the cylinders is
the necessary condition for quantum friction.

The case of the rotating body is not the only one in superfluids, where the
ergoregion is important. The ergoregion also appears for the lineraly
moving textures, where the speed of the order parameter texture exceeds the
local "speed of light"
\cite{JacobsonVolovik}.

One of us (AC) wishes to thank
the Low Temperature Laboratory of Helsinki University
of Technology for the hospitality and EU  Training and Mobility of Researches
Programme Contract N$^o$ ERBFMGECT980122 for its support.

\end{document}